# Characterization and Mitigation of Electromigration Effects in TSV-Based Power Delivery Network Enabled 3D-Stacked DRAMs

Bobby Bose
Department of Electrical & Computer Engineering
University of Kentucky, Lexington, KY, USA
bobbybose@uky.edu

Ishan Thakkar
Department of Electrical and Computer Engineering
University of Kentucky, Lexington, KY, USA
igthakkar@uky.edu

## ABSTRACT

With 3D-stacked DRAM architectures becoming more prevalent, it has become important to find ways to characterize and mitigate the adverse effects that can hinder their inherent access parallelism and throughput. One example of such adversities is the electromigration (EM) effects in the through-silicon vias (TSVs) of the power delivery network (PDN) of 3D-stacked DRAM architectures. Several prior works have addressed the effects of EM in TSVs of 3D integrated circuits. However, no prior work has addressed the effects of EM in the PDN TSVs on the performance and lifetime of 3D-stacked DRAMs. In this paper, we characterize the effects of EM in PDN TSVs on a Hybrid Memory Cube (HMC) architecture employing the conventional PDN design with clustered layout of power and ground TSVs. We then present a new PDN design with a distributed layout of power and ground TSVs and show that it can mitigate the adverse effects of EM on the HMC architecture performance without requiring additional power and ground pins. Our benchmark-driven simulation-based analysis shows that compared to the clustered PDN layout, our proposed distributed PDN layout improves the EM-affected lifetime of the HMC architecture by up to 10 years. During this useful lifetime, the HMC architecture yields up to 1.51× less energy-delay product (EDP).

## 1 Introduction

Bridging the ever-increasing performance gap between processors and DRAM requires that DRAM performance is improved even more rapidly than ever before. As a step towards meeting this requirement, 3D-stacked DRAM architectures have been developed and constantly improved over the last decade [10][11].

A 3D-stacked DRAM architecture, e.g., Hybrid Memory Cube (HMC) [9], uses through-silicon vias (TSVs) to connect multiple DRAM dies vertically in a 3D structure. Although the 3D-stacked DRAM architectures have shown significant performance improvements over the conventional 2D DRAM architectures, efforts for further improving their performance and realizing their true potential are still underway. These efforts include restructuring of 3D data organization [11][13][14][18]-[20], mitigating thermal effects [1][21]-[23], and enhancing I/O interfaces [24]-[26].

A recent work [5] has shown that the performance of 3D-stacked DRAMs is also strongly constrained by the limited IR-drop tolerance of their constituent power delivery networks (PDNs) (e.g., 75 mV IR-drop tolerance in [5]). In a 3D-stacked DRAM, the utilized PDN layout determines the worst-case PDN resistance (R), which in turn determines the peak current (I) that can be delivered by the PDN for various DRAM access operations without violating the IR-drop noise margin. In modern DRAMs, one of the basic access operations with the highest peak current requirement is subarray activation (SAA) [5][28]. Since each additional parallel SAA increases the current demand (I) from the PDN, the number of allowable parallel SAAs (NAPSAA) that can be served by a 3D-stacked DRAM without violating the IR-drop margin remains limited. This becomes the main cause that limits the performance benefits of the inherent access parallelism of 3D-stacked DRAMs, especially because the constraint on parallelism put forth by the tFAW (four bank-activates window) timing parameter is becoming irrelevant in modern DRAMs due to the diminishing dependence of the PDN on the current delivery capacity of on-chip charge pumps [5]. To this end, several prior works have focused on characterizing and mitigating the adverse effects of PDN IR-drop noise on the performance of 3D-stacked DRAM architectures [1]-[3][5]. *These works also point to the critical importance of understanding various other types of PDN-related adverse effects that can potentially harm the performance of 3D-stacked DRAMs.*

One such relevant performance-reducing effect in the PDNs of 3D-stacked DRAMs is electromigration (EM). EM typically occurs in a wire due to the mechanical displacement of metal atoms caused by the passage of electron current through the wire [8]. In a 3D-stacked DRAM, EM can cause void nucleation in the power/ground TSVs [7]. This void nucleation can grow over time, which can drastically increase the resistance of the power/ground TSVs [8] of the PDN with time. To constrict the consequent gradual increase in the IR-drop noise in the PDN below the tolerable limit, the peak deliverable current (I), and hence, NAPSAA in the 3D-stacked DRAM, must be decreased with time. However, decreasing NAPSAA reactively to address the EM-induced aging in the PDN can cause the performance of 3D-stacked DRAMs to decrease with time. Although several prior works have addressed the effects of EM in 3D ICs (e.g., [8][16][17]), to our knowledge no prior work has investigated the impacts of EM on the performance of 3D-stacked DRAMs. In this paper, *we characterize and mitigate the adverse impacts of EM-inflicted PDNs on the access parallelism, access latency, throughput, and lifetime of 3D-stacked DRAM architectures, for the first time.* We use the Hybrid Memory Cube (HMC) as an example 3D-stacked DRAM architecture for the analysis presented in this paper.

Our novel contributions in this paper are summarized below:

- We adapt a PDN layout from [3] and use it with 3D-stacked DRAMs. This new design of PDN, referred to as distributed PDN, rearranges the layout of PDN TSVs compared to the conventional clustered PDN, to reduce the IR-drop noise in the PDN, and hence, increase NAPSAA in the 3D-stacked DRAMs;
- We adopt a model from prior work that captures the effects of EM on PDN TSV resistance, and then apply this model to evaluate the EM-inflicted gradual degradation in the PDN IR-drop noise, to show that distributed PDN based 3D-stacked DRAMs achieve larger NAPSAA than clustered PDN based 3D-stacked DRAMs after being subjected to EM-induced aging;
- We perform simulation analysis for PARSEC benchmarks, to evaluate the impact of EM-inflicted decrease in NAPSAA on the access latency, throughput, and energy-delay product (EDP) of two Hybrid Memory Cube (HMC) [9] variants enabled with the clustered and distributed PDNs.

## 2 Background and Fundamentals
## 2.1 Hybrid Memory Cube (HMC)

In this paper, all our models and analysis are based on the Hybrid Memory Cube (HMC) architecture [9]. The HMC architecture typically consists of 4 to 8 DRAM die layers that are stacked on one base logic layer [9] using TSVs.

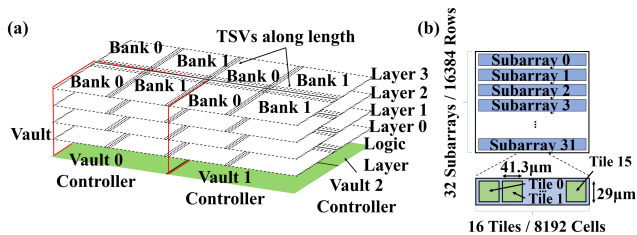

**Fig. 1: Schematic layout of (a) a 4Gb Hybrid Memory Cube (HMC) quad, and (b) a HMC bank with subarrays (SAs).**

Fig. 1 illustrates a schematic of a 4Gb HMC quad, which consists of 4 vertically structured vaults (Fig. 1(a)). Each vault spans every memory layer and has its own memory controller [9]. Every vault is split into partitions, with one partition per layer [9]. Each partition has two 128Mb banks. From Fig. 1(b), each bank consists of 16384 rows of 8Kb each. The rows in a bank are grouped into 32 subarrays vertically. Each subarray is further divided into 16 tiles of 512×512 array of cells each and 29μm×41.3μm dimensions each (22nm node).

## 2.2 3D Power Delivery Network (PDN)

In 3D integrated circuits (ICs), the power delivery network (PDN) carries power (i.e., current, voltage) from an off-chip source and delivers it to various circuit blocks on individual 3D-stacked layers. From Fig. 2(a), a 3D PDN consists of power (P) and ground (G) TSVs that pass through P/G pads on each layer. The P/G pads of each layer are connected to 2D P/G rails (Fig. 2(b)). These P/G rails form a grid through local vias (Fig. 2(b)). This grid of P/G rails delivers power to the circuit blocks on each individual layer.

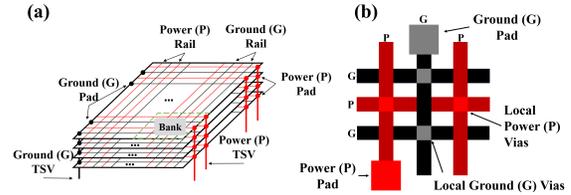

**Fig. 2: Schematic layout of TSV-based PDN for 3D ICs; Connections of power (P) and ground (G) TSVs (a) through the stack of layers, and (b) on each layer.**

In the PDNs of 3D ICs, there are two common layouts of arranging P/G TSVs: clustered and distributed layouts [3]. In the context of 3D-stacked DRAMs, the clustered layout is more commonly used [3][5]. Fig. 3(a) shows a typical clustered PDN layout from [3] adapted for the HMC architecture shown in Fig. 1. In this design, 32 P and 32 G TSVs are laid out both on the top and bottom of every DRAM bank. A rail grid of 96 vertical rails and 128 horizontal rails helps deliver power to the individual subarrays (SAs). A SA is the smallest granularity of circuit unit in a DRAM bank for which a distinct power/current consumption profile can be modeled/measured [28]. The longest current delivery path in the clustered PDN is for a SA located at the center of the bank on the top layer of the HMC stack (not shown in Fig. 3(a); more details in Section 4.1), for which the PDN exhibits the worst-case resistance ($R_W$). A typically large value of $R_W$ for the clustered PDN (e.g., $R_W \approx 0.12\Omega$ for the layout in Fig. 3(a); see more details in Section 4.1) renders relatively low peak current ($I_P$) that can be delivered without violating the IR-drop margin, exhibiting small NAPSAA for the clustered PDN enabled 3D-stacked DRAM.

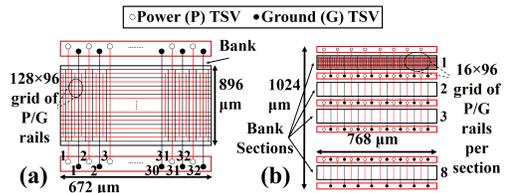

**Fig. 3: Schematic layouts of (a) clustered PDN [3], (b) proposed distributed PDN, adapted for the HMC bank from Fig. 1.**

To mitigate the adverse impacts of high PDN resistance and IR-drop in the clustered PDN layout, prior work [3] has explored the use of distributed PDN in 3D-stacked ICs. However, no prior work has investigated the use of distributed PDN for 3D-stacked DRAMs. In this paper, we adapt the distributed PDN design from [3] and use it with the HMC architecture (Fig. 3(b); see Sections 4 and 6 for detailed analysis and discussion).

## 2.3 Electromigration (EM) in TSVs

As current passes through a TSV repeatedly, a phenomenon referred to as electromigration (EM) occurs. The EM effect refers to the transfer of metal atoms due to the movement of electrons through the TSV [8]. As EM occurs in TSVs of the PDN of a 3D IC, the transfer of TSV metal atoms causes void nucleation to appear and grow in the TSVs. As a result, the resistance of the TSVs increases. For TSVs that are used for signal/data traversal in 3D ICs, it is usually agreed upon that a 10-20% increase in resistance leads to TSV failure [7][27]. However, we argue that for PDN TSVs of 3D-stacked DRAMs, it might be possible to tolerate more than 10-20% increase in their resistance, as long as the resultant increase in the PDN IR-drop noise can stay within the allowable margin (e.g., 75mV margin in [5], [15]). It would also be possible in a 3D-stacked DRAM to gradually decrease the peak current draw ($I_P$) from the PDN in response to the EM-induced increase in the TSV (and hence, PDN) resistance (R), as long as the decreased current draw still remains sufficient to support at least one subarray activation (SAA) in the DRAM. To put this into perspective, consider the clustered PDN layout from Fig. 3(a) for a DRAM bank located on the top layer of the HMC stack of Fig. 1. This design has the worst-case PDN resistance of $R_W \approx 0.12\Omega$ (Sections 2.2 and 4.1), which renders $I_P$ of ~625mA for the IR-drop margin of 0.75mV [5]. Since one SAA can draw about 100 mA current from the PDN [5], $I_P$ = 625mA can support NAPSAA of up to 6 in the DRAM bank. As a result, in response to the gradually increasing $R_W$ due to EM-induced aging, $I_P$ can be gracefully reduced from 625mA to 100mA, correspondingly decreasing NAPSAA from 6 to 1, before the EM-induced aging ends the useful lifetime of the DRAM bank, i.e., before $I_P$ and NAPSAA decrease below 100mA and 1 respectively.

## 3 Related Work

Prior research has modelled and analyzed different types of PDN structures for 3D ICs as well as 3D-stacked DRAMs. In [1], different 3D-stacked DRAM package configurations were analyzed and compared from a thermal and power delivery perspectives. In [2], various PDN models were analyzed for various 3D-stacked DRAM architectures (including HMC) that involved replacing TSVs with redistribution layers. In [3], different unit-cell topologies for TSV-based PDNs were analyzed for IR-drop noise characterization in 3D ICs. They used similar naming convention to identify various topologies (i.e., clustered and distributed). *We have transformed these single-cell topologies to their full-stack variants and used them with 3D DRAMs instead of generic 3D ICs.*

There has also been prior research into developing new PDN integrity aware request scheduling for 3D-stacked memories. In [4], a new power supply integrity conscious write scheduler is introduced to improve write throughput under power and IR-drop constraints. This paper focuses on 3D phase change memory (PCM), but the concepts can be applied to improve the performance of 3D DRAMs as well. In [5], new PDN aware scheduling and access parallelism strategies are developed to minimize IR-drop noise and improve performance in 3D-stacked DRAMs.

The effect of electromigration (EM) on TSVs is also a topic that prior research has focused on. In [6], the impact of stress on EM-induced aging of TSVs is analyzed. In [7], the effect of EM on the TSVs coupled with local vias is characterized. [8] focuses on combining the power/ground TSVs with an optimal number of local vias to minimize the routing congestion for local vias and increase the EM-inflicted lifetime of 3D ICs. *Contrary to these prior works, in this paper, we characterize how a PDN influenced by EM affects the performance and lifetime of 3D-stacked DRAMs, and show that our newly designed distributed PDN can increase the performance and lifetime of 3D-stacked DRAMs, compared to the conventional clustered PDN.*

## 4 Our Proposed Distributed PDN

As discussed in Section 2.3, the useful lifetime of a 3D-stacked DRAM (e.g., HMC) ends when EM gradually increases the worst-case PDN resistance $R_W$ beyond a threshold value from its initial value. If the initial value of $R_W$ for a PDN can be decreased, it can naturally take longer for EM to increase $R_W$ beyond the threshold, thereby increasing the useful lifetime of the DRAM. Decreasing the initial value of $R_W$ has an additional advantage of increased initial NAPSAA and related performance. Motivated by this rationale we propose to use a new PDN design with a distributed layout of P/G TSVs, instead of using the conventional PDN design with the clustered layout of P/G TSVs (Fig. 3(a)). In this section, we first describe our proposed distributed PDN. Then we present our methodology for modeling and analysis of our PDNs, before presenting our analyzed initial NAPSAA values for the HMC architecture (Fig. 1) when it is enabled with our considered PDNs.

### 4.1 Layout of Distributed PDN

Fig. 3(b) shows the schematic of our designed distributed PDN adapted for the HMC bank from Fig. 1. Compared to the clustered PDN layout depicted in Fig. 3(a), the key feature of this distributed design is that it decreases the distance the current needs to travel to reach the target subarray (SA) in the bank, without changing the total number of required P/G TSVs. Not increasing the number of TSVs has additional advantage of not increasing the required P/G pins in the package and related cost. We accomplish this by laying out the P/G TSVs throughout the bank, instead of laying them out just along the top and bottom edges of the bank (as in Fig. 3). We do this by first dividing the bank up into eight equal bank sections, with each bank section containing four subarrays (SAs). We lay the P TSVs for each bank section along its top edge, and the G TSVs along its bottom edge. To decrease the area overhead, we line up the G TSVs of one bank section with the P TSVs of another (Fig. 3(b)), which makes a maximum of only 16 TSVs being laid out in a line, compared to 32 TSVs for the clustered PDN (Fig. 3(a)). This reduces the pitch of P/G TSVs in distributed PDN. In a nutshell, the distributed PDN layout creates a separate PDN domain for each bank section, with each bank section having its dedicated P/G TSVs. This naturally decreases the worst-case current path length for the distributed PDN, rendering a lower value of initial $R_W$ and a greater NAPSAA, as discussed next.

### 4.2 Modeling of Clustered and Distributed PDNs

We model each of our considered clustered and distributed PDN designs (Fig. 3) as a 3D mesh of wire resistances, as done in [5],

using SPICE simulations. The grid of wire resistances that forms the PDN is connected to the P and G bumps on one side (not shown in Fig. 2 or 3) and is connected to subarrays (SAs) on the other side. SAs connected to the PDN are modeled as current sources which draw a fixed amount of current (100mA per SA activation (SAA) in this paper, from [5]). The values of resistances of metal wires (rails), TSVs, and bumps, along with their pitch sizes, used for modeling the PDNs are listed in Table 1.

Table 1: Parameters for modeling PDNs [5].

| Parameter | Distributed PDN | Clustered PDN |
|---|---|---|
| P/G TSV Diameter | 10 μm | |
| P/G TSV Pitch | 96 μm | 21 μm |
| P/G TSV Rail Width | 2 μm | |
| P/G TSV Rail Pitch | 8 μm | 7 μm |
| # P/G TSVs /Bank | 64 | |
| # P/G Rails /Bank | 96x128 | |
| P/G Rail Sheet Resistance | 0.9 Ω/sq | |
| TSV + C4 Resistance | 0.25 Ω | |

**Analysis of $R_W$**: To analyze the worst-case PDN resistance $R_W$, we analyze the longest current delivery path length for the clustered and distributed PDNs. The longest current delivery path occurs for the topmost layer in the HMC stack and corresponds to the central SA (SA #16 in Fig. 1(b)) in the clustered PDN, whereas it corresponds to the SA farthest from the top edge of each bank section in the distributed PDN. This longest current delivery path includes TSVs (w/ C4 bumps) through the stack (Fig. 2) along with the respective rail length. Typically, the presence of multiple parallel current channels along the delivery path (in the form of parallelly laid out rails) yields a low effective value $R_W$ (due to multiple wire resistances connected in parallel). From our SPICE simulations of the PDNs, we have evaluated $R_W$ to be 0.12Ω and 0.03Ω for the clustered and distributed PDNs respectively.

**Area Overhead**: From Fig. 3, after the addition of P/G TSVs in the distributed PDN, the width of the bank becomes 768 μm, which is 96 μm wider than the width of the bank for the clustered PDN. Similarly, the distributed PDN increases the bank height by 128 μm. Overall, the distributed PDN layout results in 1.3× larger banks in the HMC. Nevertheless, we show in Section 6 that such large area overhead can be justified, as the distributed PDN can substantially increase the performance and lifetime of the HMC.

### 4.3 Analysis of NAPSAA for PDNs

To evaluate the number of allowable parallel subarray activates (NAPSAA) for the clustered and distributed PDN enabled HMCs, we created IR-drop maps for the topmost HMC bank using our SPICE-based PDN models. To create the IR-drop maps, we swept the per-bank number of parallel SAAs through the power-of-two values of 32, 16, 8, 4, 2, and 1. Then we deem the highest value of the number of parallel SAAs, for which the IR-drop margin of 75mV is not violated in the corresponding IR-drop map, as the NAPSAA value for the respective PDN design. We choose only power-of-two values for number of parallel SAAs, as doing so simplifies the logic required to enable SA-level parallelism in HMC banks [28][29][32]. Fig. 4 shows examples of our created IR-drop maps. From the figure, the clustered PDN exhibits the highest IR-drop values of 46.9mV and 84.6mV at 4 SAAs (Fig. 4(a)) and 8 SAAs (Fig. 4(b)), respectively. Clearly, for 8 SAAs, the IR-drop margin is violated. Therefore, the clustered PDN yields NAPSAA = 4. In contrast, the distributed PDN exhibits the highest IR-drop values of 20.2mV and 64.4mV at 16 SAAs (Fig. 4(d)) and 32 SAAs (Fig. 4(c)), respectively. In Figs. 4(d) and 4(c), for 16 and 32 SAAs per bank, we had 2 and 4 SAAs per bank section, respectively. Clearly, even for 32 SAAs, the IR-drop margin is not violated, and because the HMC bank has total 32 SAs only (Fig. 1(b)), the distributed PDN yields NAPSAA = 32. We reason that, compared to the clustered PDN, the higher value of NAPSAA for the distributed PDN should render substantially better performance. It should also take longer for the EM induced aging to decrease NAPSAA below 1 (Section 2.3), yielding a longer lifetime for the distributed PDN. This reasoning of ours is corroborated by the evaluation results we present in this paper (Section 6). However, before examining our evaluation results, please consider our EM modeling methodology discussed next.

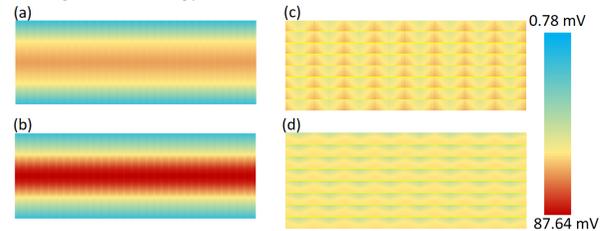

**Fig. 4: IR-drop maps for a bank on the fourth DRAM tier of the HMC (Fig. 1), for clustered PDN, with (a) 4 SAAs, (b) 8 SAAs, and for distributed PDN with (c) 32 SAAs, (d) 16 SAAs.**

## 5 Modeling of Electromigration (EM) in TSVs

In TSVs, EM is caused due to void nucleus growth [8]. To simulate void nucleus growth in the P/G TSVs of our considered PDNs, we use the void nucleation model from [7]. The rate at which the void nucleation in a P/G TSV grows depends on how much current is flowing through the TSV. This rate can be modeled in terms of vacancy flux using Eq. (1).

$$\vec{J}_V = D_V C_V \frac{eZ^*}{kT} \rho \vec{j}. \quad (1)$$

Here, $\vec{j}$ represents the current density in the P/G TSV, which changes for different levels of in-parallel SAAs. For 32, 16, 8, 4, and 2 in-parallel SAAs, we calculated $\vec{j}$ in each TSV to be 1.2e10, 6.02e9, 3.01e9, 1.5e9, 7.52e8 A/m², respectively. The vacancy concentration ($C_V$) and the vacancy diffusivity ($D_V$) in Eq. (1) can be evaluated using Eq. (2) and Eq. (3).

$$D_V = D_o \cdot \exp\left(\frac{-Ea}{kT}\right) \quad (2)$$

$$\text{and } C_V = C_o \cdot \exp\left(\frac{-Ea}{kT}\right). \quad (3)$$

We choose a time step of $dt$ = 5.0e6 sec, as it is an appropriate change in time to observe the void growth in TSVs across all values of SAAs. The void radius growth during this time step can be calculated using Eq. (4).

$$dr = \frac{\alpha f \Omega \varepsilon_{TSV} |J_v| dt}{\delta}. \quad (4)$$

We take all other variables from [6] and [7]. A compilation of parameters and values used can be seen in Table 2. From [7], we used an industrial finite element analysis (FEA) tool, Comsol multiphysics, to model the relationship between TSV void radius and resistance.

Table 2: Parameters used for modeling electromigration (EM).

| Parameter | Description | Value | Parameter | Description | Value |
|---|---|---|---|---|---|
| $\alpha$ | Ratio of captured vacancies | 1 | k | Boltzman constant (J/K) | 1.38E-23 |
| f | Ratio of vacancy volume | 0.4 | T | Temperature (K) | 453 |
| $\Omega$ | Atomic volume ($m^3$/mol) | 1.18E-29 | Z* | Charge constant (C) | 1 |
| $\delta$ | TSV void thickness (μm) | 5E-9 | $\rho_{TaN}$ | Barrier resistivity(Ωm) | 3.00E-6 |
| $D_o$ | Initial Diffusivity($m^2$/s) | 0.0047 | $\varepsilon_{TSV}$ | Effective void radius(μm) | 1.15E-6 |
| Ea | Activation Energy (V) | 1.30E-19 | $C_o$ | Atomic concentration($m^{-3}$) | 1.53E28 |

# 6 Evaluation and Discussion
## 6.1 Evaluation Setup

*6.1.1 NAPSAA Over Lifetime.* It is important to look at not only the lifetime of both PDN designs, but also what value of NAPSAA they are able to maintain in the HMC stack through aging. The NAPSAA values achievable by both PDN designs play a big role in measuring the performance of the HMC throughout its lifetime.

We used our created SPICE models for the clustered and distributed PDNs to determine how much resistance both types of PDNs could gain and still achieve certain NAPSAA. For the clustered PDN, this included NAPSAA of 4 and 2, as it has the initial NAPSAA value of 4 (Section 4.3). For the distributed PDN, this included NAPSAA of 32, 16, 8, 4, and 2, as it has the initial NAPSAA value of 32 (Section 4.3). Then, we recorded, using the EM model from Section 5, the time steps required to gain different levels of TSVs resistances for the following 9 PARSEC benchmarks: Bodytrack, Canneal, Dedup, Facesim, Ferret, Freqmine, Swaptions, Vips, and x264. To evaluate EM induced aging and its effects, we choose benchmark-driven analysis, because the inflicted amount of current density and EM stress duration in a 3D-stacked DRAM really depends on the memory access characteristics. These memory characteristics in turn depends on the running benchmark application. Using Gem5 [31], we obtained traces for the benchmark applications by running each application in the full-system mode for a short period and then capturing the region of interest. Our utilized configuration for Gem5 simulations was based on [18].

Using NVMain [30], we simulated each benchmark application on a 64Gb HMC system. Important HMC configuration parameters were taken from [29]. For every NAPSAA value for both PDN designs, the number of times (N) each application could be run till the EM induced failure occurs was recorded. This was done by dividing the time needed to gain the maximum allowed resistance, with the active time of the application (i.e., non-idle time of the application during which the memory is actively accessed by the application). Using this information, we were able to then determine at what points in the HMC's lifetime it needed to decrease NAPSAA to ensure that the EM-induced aging did not raise the PDN resistance beyond the tolerable margin.

*6.1.2 Performance Over Time.* The change in NAPSAA over time results in a change in performance as the DRAM ages. In this paper, we consider energy-delay product (EDP) for measuring performance. To analyze EDP, we ran simulations using NVMain, for a 64Gb HMC system with configuration parameters from [29]. Two copies of the HMC configuration were made, with each corresponding to a different PDN (clustered or distributed). The same 9 PARSEC benchmarks were used.

The access throughput metric was directly extracted from the results of the simulations. The average access latency, total power, and total reads and writes were also extracted directly from the NVMain output. The energy-delay product (EDP) values were calculated as {(total power / access throughput) × average latency}.

*6.1.3 Lifetime of PDNs.* During our NVMain based simulations, we also recorded the lifetime values for each HMC variant (related to the clustered and distributed PDNs) for all 9 applications. We define lifetime as the amount of time it takes for the EM induced aging to decrease NAPSAA below 1.

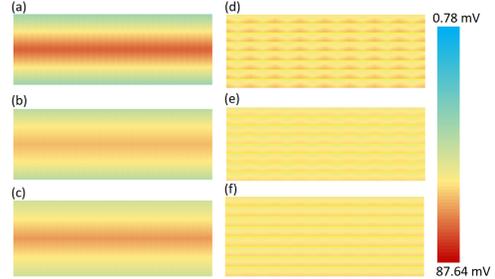

Fig. 5: EM-inflicted IR-drop maps of the HMC bank for Facesim benchmark, for the clustered PDN [(a)-(c)] and the distributed PDN [(d)-(f)] after EM aging. IR-drop maps for (a) NAPSAA=4, aging=6 years; (b) NAPSAA=2, aging=14 years; (c) NAPSAA=2, aging=40 years; (d) NAPSAA=8, aging=6 years; (e) NAPSAA=4, aging=14 years; (f) NAPSAA=2, aging=40 years.

## 6.2 Evaluation Results

*6.2.1 NAPSAA Over Lifetime.* The topmost tier of the HMC is the one that lifetime depends on and is therefore the tier we look at when analyzing NAPSAA. Fig. 5 shows the IR-drop maps for a topmost bank for different NAPSAA values related to the clustered and distributed PDNs over time, for the Facesim benchmark. From Fig. 5, after 6 and 40 years of EM-induced aging, the distributed PDN has higher NAPSAA than the clustered PDN. At 6 years of aging, the distributed PDN has NAPSAA = 8, while the clustered PDN has NAPSAA = 4 only. Similarly, at 14 years of aging, the distributed PDN has NAPSAA = 4, while the clustered PDN has NAPSAA = 2 only. At 40 years of aging, both PDN types have NAPSAA = 2 only. Nevertheless, the worst-case IR-drop in the clustered PDN is higher at about 67mV, while it is only 59.4mV in the distributed PDN. Similarly, we also analyzed NAPSAA values for all other 9 benchmark applications, but for the sake of brevity we have not included these results here. These NAPSAA values drive the performance evaluation results, discussed next.

*6.2.2 Performance Over Time.* Fig. 6 shows the energy-delay product (EDP) over time normalized to the clustered PDN enabled HMC variant's EDP at 0 years, for 9 PARSEC benchmarks. After

0, 2, 6, 10, 30 years of EM-induced aging, the distributed PDN enabled HMC achieves 1.51×, 1.24×, 1.03×, 1.20×, 1.12× better normalized EDP compared to the clustered PDN enabled HMC variant, on average across all considered PARSEC benchmarks. The higher values of NAPSAA for the distributed PDN enabled HMC yields a lower EDP for it.

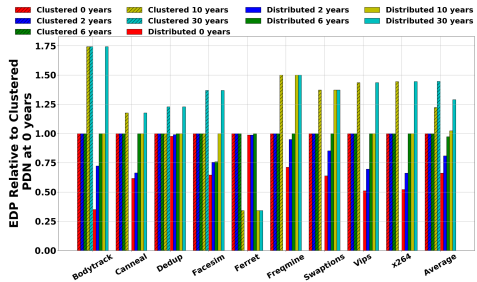

**Fig. 6: Normalized energy-delay product (EDP) over time for the clustered and distributed PDN enabled HMC variants.**

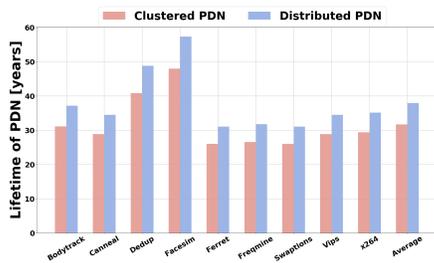

**Fig. 7: Lifetimes for the clustered and distributed PDN enabled HMC variants across 9 PARSEC benchmarks.**

*6.2.3 Lifetime of PDNs.* The lifetime values for both HMC variants for each benchmark can be seen in Fig. 7. Under the conditions of EM-induced aging, the distributed PDN enabled HMC variant achieves 1.2× higher lifetime compared to the clustered PDN on average across all considered PARSEC benchmarks, due to its reduced initial PDN resistance and improved NAPSAA values through the EM-induced aging.

## 7 Conclusion

In this paper, we introduced a new PDN design, called distributed PDN, for 3D-stacked DRAM architectures. This new PDN design reduces electromigration (EM)-induced IR-drop noise. We evaluated the EM-affected lifetime as well as performance of two Hybrid Memory Cube (HMC) variants, when they were enabled with the clustered and distributed PDNs, for 9 different PARSEC benchmark applications. We showed that our distributed PDN enabled HMC variant achieves a higher lifetime of up to 10 years. We also showed that our distributed PDN enabled HMC variant yields up to 1.51× less energy-delay product (EDP) over the HMC's useful lifetime. Thus, it can be concluded that our distributed PDN design offers a superior solution for mitigating the adverse effects of EM-induced aging on the lifetime and performance of future 3D-stacked DRAM architectures.